\documentstyle[12pt,a4]{article}
\begin{document}
\hyphenation{anti-fermion}
\topmargin= -20mm
\textheight= 230mm
\baselineskip = 0.3 in

 \begin{center}
\begin{large}
 {\bf {  VIOLATION OF  S-MATRIX FACTORIZATION  

 IN MASSIVE THIRRING MODEL }}

\end{large}

\vspace{1cm}

T. FUJITA\footnote{e-mail: fffujita@phys.cst.nihon-u.ac.jp}
 and M. HIRAMOTO\footnote{e-mail: hiramoto@phys.cst.nihon-u.ac.jp}

Department of Physics, Faculty of Science and Technology  
  
Nihon University, Tokyo, Japan

\vspace{1cm}

{\large ABSTRACT} 

\end{center}

We present a counter example which shows 
the violation of  the  S-matrix  factorization 
in  the massive Thirring model. This is done by solving the PBC 
equations of the massive Thirring model exactly but numerically. 
  The violation of the S-matrix factorization  is 
  related to the fact that the crossing symmetry  
and the factorization do not commute with each other.   This confirms  
that the soliton antisoliton S-matrix factorization  picture 
of the sine-Gordon model is semiclassical 
and does not lead to a full quantization procedure 
of the  massive Thirring model.  

\vspace{1cm}
  \par

\newpage

\begin{enumerate}
\item{\Large Introduction}

Integrable field theories in 1+1 dimensions have been studied 
in various fields of research. The Heisenberg 
 $xxx$ model is solved by the Bethe ansatz technique [1]. 
The solution is confirmed 
 most elegantly by the Yang-Baxter equation [2-4].  
Also, the Heisenberg $xxx$ model with spin $1\over 2$ is equivalent to 
the Thirring model which is solved exactly in the continuum 
field theory. Since the Yang-Baxter equation corresponds to the S-matrix 
factorization, there is, therefore, no question about 
the S-matrix factorization in this case. 

Also, the Heisenberg $xyz$ model with spin $1\over 2$ has been studied quite 
extensively [5-9]. 
Since this model is proved to be equivalent to the massive Thirring model 
in the continuum limit [10], it should be interesting to study these 
models from different points of view. 

The S-matrix factorization is also assumed 
for the massive Thirring model or sine-Gordon field theory [11]. 
With this ansatz of S-matrix factorization, 
the spectrum of the sine-Gordon model 
has been obtained and is found to be consistent with the 
spectrum obtained by the WKB method [12-14]. 
Therefore, it has been believed 
that the factorization of the S-matrix for the massive Thirring model 
holds exactly since the bound state spectrum obtained by the WKB method 
was considered to be exact. However, the recent investigations 
of the bound state problem of the massive Thirring model from the light 
cone method as well as from the Bethe ansatz technique have shown 
that there is only one bound state, on the contrary to the WKB result [15-18]. 
Indeed, it is proved that the WKB result of the bound state spectrum 
for the massive Thirring model is not exact [15-17]. This suggests that 
we should reexamine the S-matrix factorization for the 
massive Thirring model since the S-matrix factorization has also been believed 
to hold in an exact fashion. 

Until now, this factorization ansatz of the S-matrix 
for the massive Thirring model has never 
been questioned seriously. In fact, the factorization of the S-matrix 
for the particle particle scattering  holds  
exactly. This can be easily seen by using the Bethe ansatz wave functions. 
However, it is nontrivial to obtain the S-matrix 
factorization for the particle-antiparticle scattering case, as we will 
see it later. 

Now, the point is that the factorization of the S-matrix has been 
obtained for the sine-Gordon field theory with soliton anti-soliton 
pictures. Therefore, in this case, one does not have to worry about 
the antiparticle property. 
However, the soliton and antisoliton are 
only semiclassical objects and there is no clear way 
to quantize it.  This is related to the fact that the soliton is 
a solution of the field equation, but not the field itself. 

Therefore, even though  
one can easily write down the particle antiparticle scattering S-matrix 
with the factorization ansatz starting from the particle particle 
S-matrix factorization property, the identification of 
 the soliton and antisoliton as the fermion and antifermion cannot be 
 justified quantum mechanically. 

In this paper, we  present a counter example 
which shows the violation of the S-matrix factorization  
 for particle hole scattering case in the massive Thirring model. This is  
based on the Bethe ansatz solutions which we obtain exactly 
by numerically solving the PBC equations. The structure of the Bethe 
ansatz solutions is quite similar to the free fermion field theory. 
The ultraviolet cutoff $\Lambda$ is determined by the fermion number $N$ when 
we put the theory into the box with the length of $L$. In free field theory, 
the $\Lambda$ is given as
$$ \Lambda = {2\pi N\over{L}} $$
while, in the massive Thirring model, $\Lambda$ can be obtained 
as the function of the $N/L$ 
after we solve the PBC equations properly. Any physical observables 
can be obtained by letting the $N$ and $L$ infinity. 
Here, we note that this is the simplest but the best way to define 
the field theory.

In order to discuss the S-matrix factorization,  we have to determine 
the vacuum and then create particle hole states 
as long as we are interested in the charge zero sector. Since 
the number operator commutes with the hamiltonian of the massive 
Thirring model, we are only concerned with the charge zero sector. 
   
Here, we show that the S-matrix factorization does not commute 
with the crossing symmetry. Since the lagrangian of the massive 
Thirring model is invariant under the charge conjugation, one tends 
to believe that the crossing symmetry should always hold. 
Indeed, the crossing symmetry itself holds. However, one must be 
careful for the order of operations in quantum mechanics. 
There is no guarantee that one can make use of 
the crossing symmetry for the S-matrix of the particle hole 
scattering with the factorization properties together.   

In order to check the validity of the S-matrix factorization, 
we have to solve the equations of 
the periodic boundary conditions (PBC) and construct the vacuum. 
However, this is  not sufficient if we want to deal with the particle 
hole scattering. We have to solve the PBC equations for 
$n-$particle $n-$hole states. 

The important point is that the PBC equations for $n-$particle 
$n-$hole states are different not only from those for the vacuum 
states but also from each other. Therefore, we have to first solve 
these PBC equations to determine the rapidities of the vacuum 
as well as those  of the $n-$particle $n-$hole states. 

It is always an interesting question how one can construct 
the particle hole state under the situation where the rapidity 
distribution of the particle hole states are different from those 
of the vacuum state. In other words, how can we find or identify 
the rapidity of the hole state 
if the rapidities of the negative energy 
particles for the $n-$particle $n-$hole states are different 
from those of the vacuum state ? Later, we will see that we can 
find the rapidity of the hole state in the limit of infinitely  
large $N$ (the number of particle) and $L$ (the box size). 
In this limit, we can maintain the particle hole picture. 

Here, a question arises. Can we also prove the factorization 
of the S-matrix even for the particle hole scattering case ? 
As can be seen from the above statement, the hole states carry the 
information of all the negative energy particles since the rapidities 
of the particle hole state are different from the vacuum. 
Therefore, it is highly nontrivial that the S-matrix for the 
particle hole scattering can be factorized. 

Indeed, we will show below that the S-matrix for the particle 
hole scattering is not factorizable since the factorization 
and the limit of $N$ and $L$ infinity do not commute with each other. 
We stress here that, if we show any examples of the violation of 
the S-matrix factorization, this is sufficient that one cannot 
rely on the factorization ansatz. On the contrary, if we wanted to show 
the validity of the S-matrix factorization, then it would have been a 
very hard work.

The violation of the factorization ansatz indicates that
 the bound state spectrum constructed from the S-matrix 
factorization cannot be justified any more. As mentioned 
above, this is consistent with recent 
calculations on the massive Thirring model, which show 
that the Bethe ansatz PBC  equations give only one bound state, 
on the contrary to the semiclassical result and that of 
the Bethe ansatz technique with the string hypotheses. 
In fact, it is shown that there is no string-like solution 
that satisfies the PBC equations if they are solved exactly 
for the particle hole configurations [15].  
 Also, the light cone 
prescription of the massive Thirring model shows that there is only 
one bound state [16,17]. 
This indicates that 
the S-matrix factorization must correspond to the semiclassical 
approximation. This point is indeed shown in this paper, that is, 
the S-matrix factorization can be justified only when one neglects 
the commutability of the operators. This means that the soliton 
antisoliton picture is indeed semiclassical. Therefore, in order to 
make a correct correspondence between the soliton and fermion 
in a fully quantized sense, we should not rely on the S-matrix 
factorizations. 

This paper is organized as follows. In the next section, 
we will briefly describe Zamolodchikov's S-matrix factorization 
method. Then, in section 3, we 
discuss  the Bethe ansatz solution for the massive 
Thirring model. In section 4, we present the numerical solutions 
of the Bethe ansatz PBC equations in order to define the vacuum 
as well as the $n-$partcile $n-$hole states. 
Then, section 5 will treat the S-matrix for 
the particle hole scattering. In particular, we will present 
the numerical  evidences that the S-matrix of the particle hole 
scattering is not factorizable. Section 6 will summarize what we 
have understood from this work. 

\vspace{3cm}
\item{\Large Zamolodchikov's S-matrix factorization}

In integrable field theories, there are infinite conservation laws 
which induce strong selection rules [11]. 
For example, there is no particle 
creation or annihilation in the scattering process. Also, there is 
no change of the momentum before and after the scattering process. 
This means that the scattering is always elastic. 

The S-matrix for the two body 
scattering case can be defined as 
$$ S |A_i (\alpha_i) A_j(\alpha_j) >_{in} = 
\sum_{k,l} S_{ij}^{kl}(\alpha_i,\alpha_j ) 
|A_k (\alpha_i) A_l(\alpha_j) >_{out} \eqno{(2.1)} $$
where $A_i(\alpha)$'s describe the particle states with the rapidity 
$\alpha$. 

In the same way, we can consider the $N$ body S-matrix 
$$ S |A_{i_1} (\alpha_{i_1})... A_{i_N}(\alpha_{i_N}) >_{in} = 
\sum_{j_1...j_N} S_{i_1...i_N}^{j_1...j_N}(\alpha_{i_1},..., 
\alpha_{i_N} ) |A_{j_1} (\alpha_{i_1})... A_{j_N}(\alpha_{i_N}) 
>_{out} \eqno{(2.2)} $$

The basic assumption  of the S-matrix factorization is that this S-matrix 
can be written by the product of the two body S-matrices. 

For simplicity, we consider the three body case. One assumes that 
the S-matrix of the three body scattering can be written as 
$$ S_{ijk}^{lmn} = \sum_{p_1,p_2,p_3} S_{ij}^{p_1,p_2} (\alpha_{1},\alpha_{2}) 
S_{p_2k}^{p_3n} (\alpha_{3},\alpha_{1}) 
S_{p_1,p_3}^{lm} (\alpha_{2},\alpha_{3}) \eqno{(2.3)} $$
This assumption can be checked by making use of the Bethe ansatz solutions 
to the massive Thirring model for the particle-particle scattering case. 

In this three body scattering case, we can consider the following 
two different processes. In the first case, the particle 1 scatters 
with the particle 2, and then the particle 1 scatters with the particle 3, 
and finally the particle 3 scatters with the particle 2. 
On the other hand, in the second case,  the particle 2 first scatters 
with the particle 3, and then the particle 3 scatters with the particle 1, 
and finally the particle 1 scatters with the particle 2. 

Since the final state is the same between the above two processes, 
the two scattering events should give the same scattering process. Therefore, 
we obtain 
$$  \sum_{p_1,p_2,p_3} S_{ij}^{p_1,p_2} (\alpha_{1},\alpha_{2}) 
S_{p_2k}^{p_3n} (\alpha_{1},\alpha_{3}) 
S_{p_1,p_3}^{lm} (\alpha_{2},\alpha_{3}) $$
$$ = \sum_{q_1,q_2,q_3} S_{jk}^{q_2,q_3} (\alpha_{2},\alpha_{3}) 
S_{iq_2}^{lq_1} (\alpha_{1},\alpha_{3}) 
S_{q_1,q_3}^{mn} (\alpha_{1},\alpha_{2}) 
\eqno{(2.4)} $$
This is the Yang-Baxter  equation. 

Now, Zamolodchikov further assumes that the S-matrix should satisfy 
the unitarity, the analyticity and the crossing symmetry. The unitarity reads 
$$ \sum_{j_1,j_2} S_{j_1,j_2}^{i_1,i_2} 
S_{k_1,k_2}^{j_1,j_2} = \delta_{k_1,i_1} \delta_{k_2,i_2}  .
\eqno{(2.5)}  $$
The analyticity is written as 
$$ S^{\dagger}(\alpha ) = S(-\alpha^{*} )  .  \eqno{(2.6)} $$
Finally, the crossing symmetry can be stated as 
$$  S_{ij}^{kl} (\alpha ) = S_{j \bar{l}}^{\bar{i}k} (i\pi -\alpha ) 
 \eqno{(2.7)} $$
where $\bar{i}$ and $\bar{l}$ denote the anti-particles of $i$ and $l$ states. 
Here, the antisoliton of the sine-Gordon model
 is identified as the antifermion of the massive Thirring model. 
Together with the above assumptions, Zamolodchikov obtained 
the functional equation with the particles and  antiparticles included. 

However, for the massive Thirring model, 
it is nontrivial to define the antiparticle 
state in the nonperturbative sense. This is because one has to first 
determine the vacuum and then construct the particle hole states. 
The problem here is that the rapidity distributions for the vacuum 
and the particle hole states are different from each other since 
the periodic boundary condition equations are different. Therefore, 
it is nontrivial to make  hole states referring to the 
vacuum rapidity distribution. 

The question we want to address here is  
whether this crossing symmetry (eq. (2.7)) can be satisfied together with 
the factorization of the S-matrix or not. 

It is obvious that the crossing symmetry itself should hold since 
the Lagrangian of the massive Thirring model possesses this symmetry. 
Indeed, this can also be proved by the Bethe ansatz solutions. 

However, it is highly nontrivial whether the factorization for the particle 
hole scattering S-matrix can also hold or not. This is what we 
want to check in this paper and we prove numerically that the particle 
hole S-matrices do not satisfy the factorization. 

\vspace{3cm}

\item{\Large Bethe ansatz solution of massive Thirring model } 

The massive Thirring model is a 1+1 dimensional field theory with 
current current interactions [19]. Its lagrangian density can be written as 
$$  {\cal L} =  \bar \psi ( i \gamma_{\mu} \partial^{\mu} - m_0 ) \psi 
  -{1\over 2} g_0 j^{\mu} j_{\mu}   \eqno{ (3.1)} $$
where the fermion current $  j_{\mu} $  is written as
$$  j_{\mu} = :\bar \psi  \gamma_{\mu} \psi :   . \eqno{ (3.2)}   $$
Choosing a basis where $\gamma_5$ is diagonal, the hamiltonian is written as
$$  H = \int dx \left[-i(\psi_1^{\dagger}{\partial\over{\partial x}}\psi_1
-\psi_2^{\dagger}{\partial\over{\partial x}}\psi_2 )+
m_0(\psi_1^{\dagger}\psi_2+\psi_2^{\dagger}\psi_1 )+
2g_0 \psi_1^{\dagger}\psi_2^{\dagger}\psi_2\psi_1 \right]  .  \eqno{(3.3)} $$
Now, we define the number operator $N$ as 
$$ N=\int dx \psi^{\dagger}\psi   .  \eqno{(3.4)} $$
This number operator $N$ commutes with $ H$. Therefore, when we construct 
physical states, we must always consider physical quantities with the same 
 particle number $N$ as the vacuum. 
For different particle number state, the vacuum is 
different and thus the model space itself is different. 

\begin{enumerate}

\item{\large Bethe ansatz wave functions}

The hamiltonian eq.(3.3) can be diagonalized by the Bethe ansatz wave 
functions. 
The Bethe ansatz wave function $ \Psi (x_1,...,x_N)$ 
for $N$ particles can be written as [13-14] 
$$ \Psi (x_1,...,x_N)= \exp(im_0 \sum x_i \sinh \beta_i ) 
\prod_{1\leq i < j \leq N} \left[1+i\lambda (\beta_i,\beta_j) \epsilon (x_i-x_j) \right] 
\eqno{(3.5)} $$
where $\beta_i$ is related to the momentum $k_i$ 
 and the energy $E_i$ of the $i$-th particle as 
$$ k_i= m_0 \sinh \beta_i   .  \eqno{(3.6a)} $$
$$ E_i= m_0 \cosh \beta_i   .  \eqno{(3.6b)} $$
Here, $\beta_i$'s are complex variables. 

$ \epsilon (x)$ is a step function and is defined as 
$$ 
\epsilon (x) =  
\left \{
\begin{array}{rc}
-1 & x<0 \\
\ \ \\
1 & x>0 .
\end{array}
\right .
    \eqno{(3.7)} $$ 
$ \lambda (\beta_i,\beta_j)$ is related to the phase shift function 
$ \phi (\beta_i-\beta_j)$ as 
$$ \exp \left( i \phi (\beta_i-\beta_j) \right) = 
 { {1+i \lambda (\beta_i,\beta_j)}\over{1-i \lambda (\beta_i,\beta_j)}}
     .\eqno{(3.8)} $$ 
The phase shift function $\phi (\beta_i-\beta_j)$ can be explicitly written as 
$$  \phi (\beta_i-\beta_j)  =
 -2\tan^{-1} \left[{1\over 2}g_0 \tanh {1\over 2}
(\beta_i-\beta_j) \right]    . \eqno{(3.9)} $$ 
In this case, the eigenvalue equation becomes 
$$ H \mid \beta_1...\beta_N > = (\sum_{i=1}^N m_0 \cosh \beta_i ) \mid 
 \beta_1...\beta_N > \eqno{(3.10)} $$ 
where $\mid \beta_1...\beta_N >$ is related to $\Psi(x_1,...,x_N)$ as  
$$ \mid \beta_1...\beta_N >=\int dx_1...dx_N \Psi(x_1,...,x_N) 
\prod_{i=1}^N  \psi^{\dagger}(x_i,\beta_i) \mid 0 >  . \eqno{(3.11)} $$ 
Also, $\psi (x,\beta)$ can be written in terms of $\psi_1 (x)$ 
and $\psi_2 (x)$ as, 
$$ \psi (x,\beta) = e^{\beta\over 2} \psi_1(x)+ e^{-{\beta\over 2}} 
\psi_2 (x)   .  \eqno{(3.12)}     $$ 
From the definition of the rapidity variable $\beta_i$'s, one sees that 
for positive energy particles, $\beta_i$'s are real while for 
negative energy particles, $\beta_i$ takes the form $i\pi -\alpha_i$ 
where $\alpha_i$'s are real. Therefore, in what follows, we denote 
the positive energy particle rapidity by $\beta_i$ and the negative 
energy particle rapidity by $\alpha_i$. 

\vspace{1cm}
\item{\large Periodic Boundary Conditions }

The Bethe ansatz wave functions satisfy the eigenvalue equation [eq.(3.10)]. 
However, they still do not have proper boundary conditions. The 
simplest way to define field theoretical models is to put the theory in a 
box of length $L$ and impose periodic boundary conditions (PBC) on the states. 

Therefore, we demand that $\Psi (x_1,..,x_N)$ be periodic in each argument 
$x_i$. This gives the boundary condition 
$$ \Psi (x_i=0) =  \Psi (x_i = L)   . \eqno{(3.13)} $$ 
This leads to the following PBC equations, 
$$ \exp (im_0 L \sinh \beta_i )= \exp (-i \sum_j \phi (\beta_i -\beta_j) ) 
   .   \eqno{(3.14)} $$ 
Taking the logarithm of eq.(3.14), we obtain 
$$ m_0 L \sinh \beta_i = {2\pi n_i} - \sum_j \phi (\beta_i -\beta_j)  
      \eqno{(3.15)} $$ 
where $n_i$'s are integers. These are equations which we should now solve. 

\end{enumerate}

\vspace{3cm}
\item{\Large Numerical Solutions}

The parameters we have in the PBC equations  
are the box length $L$ and the particle number $N$. In field theory, 
we  introduce the ultraviolet cutoff parameter. In the PBC equations, 
the cutoff parameter is the particle number $N$. This is quite similar 
to the free field theory where the cutoff 
parameter is defined by the particle number $N$. Once we determine the 
particle number $N$ and the box length $L$, then we determine 
all the rapidity values necessary to obtain any physical quantity. 
In this sense, the Bethe ansatz solution can be well defined as a field 
theory. Any physical observables can be obtained by letting the $N$ 
and $L$ infinity. 

There is also an important quantity which is 
the density of the system $\rho$. It becomes  
$$ \rho = {N \over{L}}    .  \eqno{(4.1a)}  $$ 
Here, the system is fully characterized by
the density $\rho$. For later convenience, we define the effective 
density $\rho_0$ as   
$$ \rho_0 = {N_0\over{ L_0}}  \eqno{(4.1b)} $$ 
where  $L_0$ and $N_0$ are defined as $L_0 = m_0 L$ and 
 $N_0 = {1\over 2}(N-1)$, respectively.

It is important to note that the beta function of the massive Thirring model 
 vanishes to all orders [20], and thus there is no need to 
worry about the renormalization of the coupling constant. However, one has 
to be careful for the coupling constant normalization arising from 
the way one makes the current regularization.  Here, we employ 
the Schwinger's normalization throughout this paper. 

In what follows, we  solve the PBC equations numerically 
in the same manner as in ref.[15]. The numerical method to solve 
the PBC equations is explained in detail in ref.[15]. We note that 
the errors involved in the numerical calculations are very small. 
For example, the numerical uncertainty for the rapidity values 
 will appear  at the  $10^{-6}$ level, and therefore, this does not 
 generate any problem in the present discussion.

\vspace{1cm}

\begin{enumerate}
\item{\large Vacuum state}

The PBC equations 
for the vacuum which is filled with negative energy particles 
( $\beta_i=i\pi -\alpha_i$ ),  become
$$  \sinh \alpha_i = {2\pi n_i \over{L_0}}  
 - {2\over{L_0}} \sum_{j \not= i} \tan^{-1}\left[{1\over 2}g_0 
\tanh{1\over 2} (\alpha_i -\alpha_j) \right] ,  \qquad  
(i=1,..,N)    \eqno{(4.2)} $$ 
where $n_i$ runs as 
$$ n_i = 0, \pm 1, \pm 2, ..., \pm N_0    .  $$

There is no ambiguity to determine the vacuum. The vacuum is indeed 
constructed uniquely [15]. 

\vspace{1cm}
\item{\large $1p-1h$ configuration}
 
One particle-one hole $(1p-1h)$ 
 state can be made by taking out one 
negative energy particle (the $n_{i_0}$ particle) 
 and putting it into a positive energy state. 
In this case, the PBC equations become 
\renewcommand{\theequation}{4.3\alph{equation}}
\setcounter{equation}{0}
\begin{eqnarray}
n_i\neq n_{i_0} \nonumber \\
 & \sinh\alpha_i & = \frac{2\pi n_i}{L_0}-\frac{2}{L_0}\tan^{-1}
 \left[ {1\over 2}g_0\coth\frac{1}{2}(\alpha_i+\beta_{i_0})\right] 
 \nonumber \\
 & &- \frac{2}{L_0}\sum_{j\neq i,i_0}\tan^{-1}
\left[{1\over 2}g_0\tanh\frac{1}{2}
 (\alpha_i-\alpha_j)\right] \\
 \nonumber \\
n_i=n_{i_0} \nonumber \\
 & \sinh\beta_{i_0} & = \frac{2\pi n_{i_0}}{L_0}+\frac{2}{L_0}\sum_{j\neq i_0}
 \tan^{-1}\left[{1\over 2}g_0\coth\frac{1}{2}(\beta_{i_0}+\alpha_j)\right]   
\end{eqnarray} 
where $\beta_{i_0}$ can be a complex variable as long as it can satisfy 
eqs.(4.3). 

It is important to note that the momentum allowed for the 
positive energy state must be determined by the PBC equations. Also, 
the momenta occupied by the negative energy particles  are different from 
the vacuum case as long as we keep the values of $N$ and $L$ finite, 
as can be seen from eqs.(4.2) and (4.3). 

Note that the lowest configuration one can consider 
is the case in which one takes 
out $n_i=0$ particle and puts it into the positive energy state. 
 This must be the first excited state since it has a  symmetry of 
$\alpha_i = - \alpha_{-i}$. Indeed, as discussed in ref.[15], 
this state corresponds to the only bound state (the boson state) 
in this model. 

Next, we consider the following configurations 
in which we take out $n_i= \pm 1, \pm  2 ,..$ particles 
and put them into the positive energy state.  
These configurations of one particle-one hole 
states turn out to be the scattering states [15].

\vspace{1cm}
\item{\large $2p-2h$ configurations}

In the same way as above, we can make two particle-two hole  $(2p-2h)$ states. 
Here, we take out the $n_{i_1}$ and  $n_{i_2}$ particles 
and put them into positive energy states.  
The PBC equations for the two particle-two hole states become  
\renewcommand{\theequation}{4.4\alph{equation}}
\setcounter{equation}{0}
\begin{eqnarray}
n_i\neq n_{i_1}, n_{i_2} \nonumber \\
 & \sinh\alpha_i & =  \frac{2\pi n_i}{L_0}-\frac{2}{L_0}\tan^{-1}
 \left[ {1\over 2}g_0\coth\frac{1}{2}(\alpha_i+\beta_{i_1})\right] 
 \nonumber \\
& & -\frac{2}{L_0}\tan^{-1}
 \left[ {1\over 2}g_0\coth\frac{1}{2}(\alpha_i+\beta_{i_2})\right] 
 \nonumber \\
 & & - \frac{2}{L_0}\sum_{j\neq i,i_1,i_2}\tan^{-1}
\left[{1\over 2}g_0\tanh\frac{1}{2}
 (\alpha_i-\alpha_j)\right] \\
 \nonumber \\
 n_i=n_{i_1} \ \ \ \nonumber \\
 & \sinh\beta_{i_1} & = \frac{2\pi n_{i_1}}{L_0} +\frac{2}{L_0}\tan^{-1}
\left[{1\over 2}g_0\tanh\frac{1}{2}
 (\beta_{i_1}-\beta_{i_2})\right] \nonumber \\
&&+\frac{2}{L_0}\sum_{j\neq i_1,i_2}
 \tan^{-1}\left[{1\over 2}g_0\coth\frac{1}{2}(\beta_{i_1}+\alpha_j)\right] \\ 
\nonumber \\
n_i=n_{i_2} \ \ \ \nonumber \\
 & \sinh\beta_{i_2} & = \frac{2\pi n_{i_2}}{L_0} + \frac{2}{L_0}\tan^{-1}
\left[{1\over 2}g_0\tanh\frac{1}{2}
 (\beta_{i_2}-\beta_{i_1})\right] \nonumber \\
&&+\frac{2}{L_0}\sum_{j\neq i_1,i_2}
 \tan^{-1}\left[{1\over 2}g_0\coth\frac{1}{2}(\beta_{i_2}+\alpha_j)\right] .  
\end{eqnarray} 

All the configurations one can construct for the two particle two hole 
states turn out to be the scattering states 
in this model [15].

\end{enumerate}

\vspace{3cm}
\item{\Large Factorization of S-matrix}

Since  the vacuum state is denoted  by the rapidities $\alpha_1,..,\alpha_N$,  
$$ | {\rm vac}> = |\alpha_1, ... ,\alpha_N >   $$
we can  define the S-matrix of the vacuum to vacuum transition. In this case,  
we can write it as
$$  <{\rm vac}| S | {\rm vac} > = \prod_{i>j}^{N}S_0(\alpha_i, \alpha_j ) 
\eqno{(5.1)} $$ 
where $ S_0(\alpha_i, \alpha_j ) $ denotes the two body S-matrix 
between $i$ and $j$ particles to make a transition from $x_i < x_j$ 
to $x_i > x_j $ states. It can be written as 
$$  S_0(\alpha_i,\alpha_j) 
= \exp \left( i \phi (\alpha_i- \alpha_j ) \right) $$
where $  \phi (\alpha_i-\alpha_j) $ can be written explicitly as   
$$    \phi (\alpha_i-\alpha_j) =   
   2\tan^{-1} \left[{1\over 2}g_0 \tanh {1\over 2}
(\alpha_i-\alpha_j) \right]    . \eqno{(5.2)} $$ 
Note that these particles here 
are negative energy ones. Obviously, the vacuum to vacuum 
transition must be unity, 
$$ <{\rm vac} | S | {\rm vac}> = 1  . \eqno{(5.3)} $$
Therefore, we have always the constraint 
$$ \prod_{i>j}^{N} S_0 ( \alpha_i,\alpha_j) = 1  . \eqno{(5.4)} $$

We also give a correspondence between 
$ S_0(\alpha_i, \alpha_j )$ as defined here 
and $S_{pp}^{pp}(\alpha_i,\alpha_j)$ as defined in section 2.  
$$  S_{pp}^{pp} (\alpha_i , \alpha_j ) =  S_0(\alpha_i, \alpha_j ) 
\eqno{(5.5)} $$
where $pp$ denotes the particle particle scattering. If it is for 
particle hole scattering case, the S-matrix can be written as 
$ S_{ph}^{hp} (\alpha_i,\beta_j^h ) $ 
 as will be soon explained below. 
It should be noted that, in the massive Thirring model, the states 
are completely specified by the rapidity variables $ \alpha $ with the index 
of particle or hole. These are the quantum numbers that 
can specify the states.

\vspace{1cm}
\begin{enumerate}
\item{\large S-matrix}

Now, we first define  the one particle one hole state. 
This can be denoted as 
$$ |1p1h> = |\beta_1, \alpha^{\dagger}_2,..,\alpha^{\dagger}_N > 
\equiv |\beta_1,\beta_1^{h} > \eqno{(5.6)} $$ 
where $\beta_1^h$ is the rapidity of the hole state when 
$N$ and $L$ become infinity. But for the moment, it is only 
symbolically written since $\alpha_i^{\dagger}$'s differ from 
 $\alpha_i$'s of the vacuum state. 

Therefore, the S-matrix for the $1p-1h$ state can be written as
$$ S_{1p1h} (\beta_1,\alpha^{\dagger}_2,..,\alpha^{\dagger}_N) 
= \prod_{i>j> 1}^{N}S_0(\alpha^{\dagger}_i, \alpha^{\dagger}_j )
\prod_{k=2}^{N}S_0(\beta_1, \alpha^{\dagger}_k )  . \eqno{(5.7)} $$ 
Here, we note that, in the large $N$ and $L$ limits, we find that 
$$ \alpha^{\dagger}_i \rightarrow \alpha_i    $$ 
as will be shown later numerically in Table 1. 

In what follows, we only discuss those particle hole 
scattering processes where 
the particle states are always involved. We do not consider here 
particle particle scattering or hole hole scattering processes. 

We note that this particular example is sufficient since 
we are only interested in finding an example of the violation 
of the factorization. 

Therefore, we define the S-matrix for the particle-hole scattering 
in the following way, 
$$ S_{ph}^{hp} (\beta_1,\beta_1^h ) \equiv 
\prod_{k=2}^{N}S_0(\beta_1, \alpha^{\dagger}_k )  . \eqno{(5.8)} $$
Next, we want to define  the two particle two hole state. 
This can be denoted as 
$$ |2p2h> = |\beta_1, \beta_2, \alpha^{\ddagger}_3,..,\alpha^{\ddagger}_N > 
\equiv |\beta_1,\beta_1^{h} ,\beta_2,\beta_2^{h} >   . 
\eqno{(5.9)} $$
Again, $\beta_1^h$ and $\beta_2^h$ are written symbolically. 

Therefore, the S-matrix for the $2p-2p$ state can be written as
$$ S_{2p2h} (\beta_1,\beta_2,\alpha^{\ddagger}_3,..,\alpha^{\ddagger}_N) $$
$$= S_0(\beta_1,\beta_2)
 \prod_{i>j>2}^{N}S_0(\alpha^{\ddagger}_i, \alpha^{\ddagger}_j )
\prod_{k=3}^{N}S_0(\beta_1, \alpha^{\ddagger}_k )
\prod_{l=3}^{N}S_0(\beta_2, \alpha^{\ddagger}_l )  . \eqno{(5.10)}  $$ 
Here again, we note that, in the large $N$ and $L$ limits, we find that 
$$ \alpha^{\ddagger}_i \rightarrow \alpha_i  .   $$ 
In the same way as $S_{ph}^{hp}$, we define the $2p-2h$ scattering 
process where the particle states are involved. 
$$ S_{phph}^{hphp} (\beta_1,\beta_1^h,\beta_2,\beta_2^h ) \equiv
\prod_{k=3}^{N}S_0(\beta_1, \alpha^{\ddagger}_k )
\prod_{l=3}^{N}S_0(\beta_2, \alpha^{\ddagger}_l )  . \eqno{(5.11)}  $$ 
Now that the  $ \alpha^{\dagger}_i $ and $ \alpha^{\ddagger}_i $ approach 
to the $ \alpha_i $ for the large $N$ and $L$ limits, 
one may think that one can replace 
the  $ \alpha^{\dagger}_i $ and $ \alpha^{\ddagger}_i $ by the $\alpha_i$. 
We denote these S-matrices by $ S_{ph}^{hp} (\beta_1,\beta_1^h )_{(V)} $ and 
$ S_{phph}^{hphp} (\beta_1,\beta_1^h,\beta_2,\beta_2^h )_{(V)} $. 
Therefore, if we take the large $N$ and $L$ limits at this stage 
without care for the difference between $\alpha_i $, 
$  \alpha^{\dagger}_i $ and $  \alpha^{\ddagger}_i $, then 
we obtain that the S-matrix 
$ S_{phph}^{hphp} (\beta_1,\beta_1^h,\beta_2,\beta_2^h )_{(V)} $ 
for the $2p-2h$ state  
can be factorized into $1p-1h$ S-matrices. That is, 
$$ S_{phph}^{hphp} (\beta_1,\beta_1^h,\beta_2,\beta_2^h )_{(V)} = 
S_{ph}^{hp} (\beta_1,\beta_1^h )_{(V)} S_{ph}^{hp} (\beta_2,\beta_2^h )_{(V)}  . \eqno{(5.12)} $$

Eq.(5.12) shows that the S-matrix of the two$-$particle 
two$-$hole scattering seems indeed factorized. 
This is the main reason why people believe that the factorization 
ansatz for the particle hole scattering should hold in the same 
way as the massless case or in other words as the particle 
particle scattering case. 

Based on the factorization ansatz of the S-matrix, 
Zamolodchikov and Zamolodchikov obtained 
the Yang-Baxter type functional equation for the 
S-matrix which can determine the shape of the S-matrix itself 
by the simple algebraic manner [11]. However, the derivation of the functional 
equation implicitly assumes  the factorization of the particle hole 
scattering S-matrix.  Even though the factorization of the particle 
hole S-matrix is valid in terms of their rapidity variables, it is 
nontrivial to show that this factorization can hold even for the common 
rapidity variables with the vacuum.  
 
Therefore, one must be  careful for the treatment of the rapidity variables 
when one calculates the S-matrix. In particular, one should carefully 
treat the order of the operations when one makes the large $N$ and $L$ 
limit in quantum mechanics. The large limits of $N$ and $L$ should be 
taken as late as possible. 

\vspace{1cm}

\item{\large Large $L$ and $N$ limits }
 
In what follows, we show that, in the large $N$ and $L$ limits,  eq.(5.12) 
does not hold.  
In order to see it, we first rewrite the  S-matrix of the $1p-1h$ state  
in terms of the vacuum rapidities. 
$$  S_{ph}^{hp} (\beta_1,\beta_1^h )  
= \prod_{k=2}^{N}S_0(\beta_1, \alpha_k+\delta_k ) \eqno{(5.13)} $$ 
where $\delta_i$ is defined as the difference between $\alpha_i$ 
and $\alpha_i^{\dagger}$, that is, $\delta_i =\alpha_i^{\dagger} 
-\alpha_i$ . 

Since $\delta_i$ is quite small for the large values of $N$ and $L$, 
we can expand 
$ S_0(\beta_1, \alpha_j+\delta_j ) $ in terms of $\delta_i$.  
Here, we expand the phase shift function 
$ \phi (\beta_1+ \alpha_j+\delta_j-i\pi ) $ 
$$ S_0(\beta_1, \alpha_j+\delta_j ) = \exp \left[ 
i\phi (\beta_1+ \alpha_j-i\pi) ) 
+ i \delta_j {\partial\over{\partial \alpha_j}} 
 \phi (\beta_1+ \alpha_j-i\pi )  \right.  $$ 
$$ \left.  + {i\over 2} \delta_j^2 {\partial^2\over{\partial \alpha_j^2}} 
 \phi (\beta_1+ \alpha_j-i\pi ) + ...\right]  \eqno{(5.14)} $$ 
The first terms of eqs.(5.14) correspond to the S-matrices which 
are factorizable. 

Now, the important point is that the $\delta_i$ itself becomes 
zero as the values of $N$ and $L$ become infinity but 
the summation of $\delta_i$ stays finite in this limit. 
We will discuss it in detail below. 
Therefore, it is important to keep 
the second term  for the S-matrix evaluation. 

In the same way, we rewrite the S-matrix of the $2p-2h$ state.  
$$ S_{phph}^{hphp} (\beta_1,\beta_1^h,\beta_2,\beta_2^h ) 
= 
\prod_{k=3}^{N}S_0(\beta_1, \alpha_k+\epsilon_k )
\prod_{l=3}^{N}S_0(\beta_2, \alpha_l+\epsilon_l ) \eqno{(5.15)}  $$ 
where $\epsilon_i$ is defined as 
 $\epsilon_i=\alpha_i^{\ddagger} -\alpha_i $. Since $\epsilon_i$ is 
very small for the large values of $N$ and $L$, we can expand 
$ S_0(\beta_1, \alpha_k+\epsilon_k ) $ and 
$ S_0(\beta_2, \alpha_l+\epsilon_l )$  
in terms of $\epsilon_i$ in the same way as eqs.(5.14) 
$$ S_0(\beta_1, \alpha_k+\epsilon_k ) = \exp 
\left[i \phi (\beta_1+ \alpha_k-i\pi ) 
+ i\epsilon_k {\partial\over{\partial \alpha_k}} 
 \phi (\beta_1+ \alpha_k-i\pi )  \right. $$    
$$ \left.  + {i\over 2}  \epsilon_k^2 {\partial^2\over{\partial \alpha_k^2}} 
 \phi (\beta_1+ \alpha_k-i\pi ) + ...\right]  \eqno{(5.16a)} $$ 
$$ S_0(\beta_2, \alpha_l +\epsilon_l) = \exp \left[ 
i\phi (\beta_2+ \alpha_l-i\pi ) 
+i \epsilon_l {\partial\over{\partial \alpha_l}} 
 \phi (\beta_2+ \alpha_l-i\pi )  \right. $$   
$$ \left. + {i\over 2}\epsilon_l^2 {\partial^2\over{\partial \alpha_l^2}} 
 \phi (\beta_2+ \alpha_l-i\pi ) + ...\right]  \eqno{(5.16b)} $$ 

We can also check later that the $\epsilon_i$ itself becomes zero 
when $N$ and $L$ become infinity. However, the summation of 
$\epsilon_i$ is finite and therefore, we have to keep them when 
we want to show the S-matrix factorization. 

Now, we rewrite eqs.(5.13) and (5.15) in terms of 
eqs.(5.14) and (5.16). 
As mentioned above,  we are only concerned 
with the particle hole scattering $S$ matrix factorization. 

Therefore, eqs.(5.13) and (5.15) can be written as 
$$  S_{ph}^{hp} (\beta_1,\beta_1^h )  
=   \exp \left[ i\sum_{m=2}^{N} \phi (\beta_1+ \alpha_m-i\pi ) \right]  $$
$$ \times \exp \left[ i \sum_{k=2}^{N} \left( \delta_k  
  {\partial \phi (\beta_1+\alpha_k-i\pi) \over{\partial \alpha_k}}  
    + {1\over 2}\delta_k^2   {\partial^2 \phi (\beta_1+\alpha_k-i\pi)
   \over{\partial \alpha_k^2}} 
        +... \right)    \right] .  \eqno{(5.17)} $$ 

$$ S_{phph}^{hphp}(\beta_1,\beta_1^h,\beta_2,\beta_2^h) = 
 \exp \left[ i\sum_{m=3}^{N} 
\phi (\beta_1+ \alpha_m-i\pi ) + i\sum_{m=3}^{N} \phi (\beta_2+ \alpha_m-i\pi )
  \right]  $$
$$ \times \exp \left[ i \sum_{k=3}^{N} \left( 
 \epsilon_k {\partial \phi (\beta_1+\alpha_k-i\pi) \over{\partial \alpha_k}} 
   +  {1\over 2} \epsilon_k^2   {\partial^2 \phi (\beta_1+\alpha_k-i\pi) 
  \over{\partial \alpha_k^2}}  \right)
     \right.  $$  
$$   + \left. i \sum_{k=3}^{N} \left( 
 \epsilon_k {\partial \phi (\beta_2+\alpha_k-i\pi) \over{\partial \alpha_k}} 
   +  {1\over 2} \epsilon_k^2  {\partial^2 \phi (\beta_2+\alpha_k-i\pi) 
  \over{\partial \alpha_k^2}}   \right)   
   +...  \right]   .   \eqno{(5.18)} $$

Clearly from eqs.(5.17) and (5.18), the S-matrix factorization 
cannot hold unless the exponential 
terms except the first one vanish all together. 

The important point is that $\sum \delta_k $ contains all the information 
of the $1p-1h$ state and therefore depends on $\beta_1$ and $\beta_1^h$ 
which are the quantum numbers that characterize the $1p-1h$ state. In the 
same manner, the summation $\sum \epsilon_k$ depends on 
$\beta_1$, $\beta_1^h$, $\beta_2$ and $\beta_2^h$ which characterize 
the $2p-2h$ state. 

\vspace{1cm}
\item{\large Numerical results}

Now, we want to present the numerical check of the various 
quantities which appear in eqs. (5.17) and (5.18). 
Here, we first solve numerically the PBC equations (4.2), 
(4.3) and (4.4). We vary the number of particle $N$ and the box 
size $L_0$. 

First, we want to check how  the rapidity 
values of the $1p-1h$ and  $2p-2h$ configurations 
converge into those of the vacuum 
state when $N$ becomes large. In Table 1, we plot several values 
of the rapidity as the function of $N$ with the fixed value of the 
density $\rho_0$. This clearly shows that 
the rapidities of the $1p-1h$ and $2p-2h$ 
states approach to those of the vacuum state 
when $N$ becomes very large. Therefore, eq.(5.12) 
indeed holds. However, we must be careful when we treat any physical 
quantities which depend on the sum of the rapidity difference 
between the vacuum state and the $1p-1h$ or $2p-2h$ configurations.  
The important point is that, even though each rapidity of 
the $1p-1h$ and $2p-2h$ 
states approaches to that of the vacuum, the sum of the rapidity 
differences stays finite. 

In order to see the effect mentioned above, 
we define the following quantities
$$ D_{1p1h}(n_1) \equiv D_{1p1h}(\beta_1,\beta_1^h) =  
 \sum_{k=2}^{N}  \delta_k 
{\partial \phi (\beta_1+\alpha_k-i\pi) 
 \over{\partial \alpha_k}}  \eqno{(5.19a)} $$
$$ E_{1p1h}(n_1) \equiv E_{1p1h}(\beta_1,\beta_1^h)= {1\over 2} \sum_{k=2}^N  
     \delta_k^2   {\partial^2 \phi (\beta_1+\alpha_k-i\pi)
   \over{\partial \alpha_k^2}}     \eqno{(5.19b)} $$
$$ D_{2p2h}(n_1,n_2) \equiv D_{2p2h}(\beta_1,\beta_2, \beta_1^h, \beta_2^h) 
=  \sum_{k=3}^{N} \epsilon_k 
{\partial \phi (\beta_1+\alpha_k-i\pi) 
 \over{\partial \alpha_k}}  \eqno{(5.19c)} $$
$$ E_{2p2h}(n_1,n_2) \equiv E_{2p2h}(\beta_1,\beta_2,\beta_1^h,\beta_2^h) 
= {1\over 2} \sum_{k=3}^N  
     \epsilon_k^2  {\partial^2 \phi (\beta_1+\alpha_k-i\pi)
   \over{\partial \alpha_k^2}}    \eqno{(5.19d)} $$
where $\beta_1$ and $\beta_2$ correspond to the rapidity values for 
the $n_1$ and $n_2$ states. 

Here $ {\partial \phi (\beta_1+\alpha_k-i\pi) 
 \over{\partial \alpha_k}} $ and 
$ {\partial^2 \phi (\beta_1+\alpha_k-i\pi)  \over{\partial \alpha_k^2 }} $ 
can be written explicitly using eq.(3.9) as
$$ {\partial \phi (\beta_1+\alpha_k-i\pi)  \over{\partial \alpha_k}} 
 = {g_0\over 2} {1\over{ {g_0^2\over 4} + (1+ {g_0^2\over 4}) 
 \sinh ^2 {1\over 2}(\beta_1 + \alpha_k ) }}  .  \eqno{(5.20)} $$
$$ {\partial^2 \phi (\beta_1+\alpha_k-i\pi)  \over{\partial \alpha_k^2 }} 
 =- {g_0\over 2} {(1+{g_0^2\over 4}) 
 \sinh {1\over 2}( \beta_1 + \alpha_k ) 
 \cosh {1\over 2}( \beta_1 + \alpha_k ) 
  \over{ \left( {g_0^2\over 4} + (1+ {g_0^2\over 4}) 
 \sinh ^2 {1\over 2}(\beta_1 + \alpha_k ) \right)^2 }}  .  
 \eqno{(5.21)} $$

Since we have solved the PBC equations numerically, we know all the 
rapidity values, and thus we can calculate eqs.(5.19) together with 
eq.(5.20) and eq.(5.21).

In Table 2, we plot the values of $D_{1p1h}$, $E_{1p1h}$, 
$D_{2p2h}$ and $E_{2p2h}$  
as the function of $N$ and $L_0$ for the lowest scattering state. 
As can be seen, the calculated values 
of $D_{1p1h}$ and $D_{2p2h}$ do not depend very much 
on the $N$ and $L_0$ as long as we keep the same density $\rho_0 $. 

However, as can be seen from Table 2, 
the $E_{1p1h}$ and the $E_{2p2h}$ decrease as $N$ 
increases. They behave as 
$$ E_{1p1h}(n_1) \sim {1\over N} $$
$$ E_{2p2h}(n_1,n_2) \sim {1\over N}   . $$
Therefore, these quantities do not survive at the large $N$ limit. 

This clearly shows that the sum of the rapidities times the derivative 
of the S-matrix 
is finite even though each rapidity value becomes zero when $N$ and $L$ 
become infinity. Therefore, we prove that these second 
terms do not vanish. 

In order to see more clearly the above effects, we rewrite 
the $S_{ph}^{hp}(\beta_1,\beta_1^h)$ and 
$S_{phph}^{hphp}(\beta_1,\beta_1^h, \beta_2,\beta_2^h)$ 
in terms of $D_{1p1h}(n_1)$ and 
$D_{2p2h}(n_1,n_2)$. 
$$ S_{ph}^{hp}(\beta_1,\beta_1^h)=S_{ph}^{hp}(\beta_1,\beta_1^h)_{(V)} 
\exp \left[iD_{1p1h}(n_1) \right] \eqno{(5.22)} $$ 
$$ S_{phph}^{hphp}(\beta_1,\beta_1^h,\beta_2,\beta_2^h)
 =  S_{ph}^{hp} (\beta_1,\beta_1^h )_{(V)} 
S_{ph}^{hp} (\beta_2,\beta_2^h )_{(V)} $$
$$ \times \exp \left[iD_{2p2h}(n_1,n_2) + 
iD_{2p2h}(n_2,n_1) \right] \eqno{(5.23)} $$ 
Now, it is obvious that the  
$D_{1p1h}(n_1)$ and 
$D_{2p2h}(n_1,n_2)$ 
are quite different from each other
since they carry the information of the many body nature 
of the particle hole states. Indeed, the 
$D_{2p2h}(n_1,n_2)$ 
depends on $\beta_1$, $\beta_2$,  $\beta_1^h$ and 
$\beta_2^h$. Therefore,  
the factorization of the particle hole S-matrix cannot be justified. In fact, 
as we see below, the numerical results confirm the difference 
between the $D_{1p1h}(n_1)$ and 
$D_{2p2h}(n_1,n_2)$ . 

\vspace{1cm}
\item{\large Field theory limit}

Now, we should take the field theory limit since this is the only way to 
compare our results with Zamolodchikov's factorization ansatz. For the 
field theory limit, we have to let 
$\rho \rightarrow \infty$ [15]. In the case of the bound state problem, 
when we let $\rho \rightarrow \infty$, then 
we should take the limit of $m_0 \rightarrow 0$, keeping 
the excited state energy finite. In order to see it more concretely, 
we repeat here how it is done for the bound state problem.  
The excited state energy $\Delta E$ is written as [15] 
$$ \Delta E = m_0 \left( A+B \left( {\rho\over{m_0}} \right)^{\alpha} 
\right) $$ 
where $A$ and $B$ are some constants which depend on the coupling constant. 
Here, $\alpha$ is a constant with positive value  which is smaller 
than unity. In this case, when we make $\rho \rightarrow 
\infty$, keeping $\Delta E$ 
finite, we can make a fine tuning of $m_0$ such that 
$$  m_0^{1-\alpha} \rho^{\alpha} = {\rm finite}  . $$ 
This means that we should let $m_0 \rightarrow 0 $. 

However, in the present case, 
 the situation is a bit different 
and even simpler  
since the correction factor itself does not have any dimensions. 
Therefore, the field theory limit is just that we let the $\rho_0$ infinity. 
In fig.1, we show the  calculated absolute values of $D_{1p1h}$ and $D_{2p2h}$ 
as the function of $\rho_0^{-1}$. 
 In this case, they  
 can be well parameterized by the following 
functions as 
$$ D_{1p1h}(n_1) = C^{(1)} {\rho_0}^{-1} +D^{(1)} 
\eqno{(5.24a)} $$  
$$ D_{2p2h}(n_1,n_2) = C^{(2)} \exp \left(-{\kappa_2\over{\rho_0}} \right) 
+D^{(2)} \eqno{(5.24b)} $$  
where  $C^{(i)}$, $D^{(i)}$ and $\kappa_2$ are some constants 
which depend on the coupling constant and the rapidity values of 
the particle  hole states. 

As can be seen, when the $\rho_0$ 
goes to infinity, then the $D_{1p1h}$ and $D_{2p2h}$ approach to 
some finite numbers. Namely, they become
$$ D_{1p1h}(n_1) \rightarrow D^{(1)} \eqno{(5.25a)} $$  
$$ D_{2p2h}(n_1,n_2) \rightarrow C^{(2)} +D^{(2)} \eqno{(5.25b)} $$  

This means that we have made the field theory limit 
in a correct way.  The values of $C^{(i)}$, $D^{(i)}$ and $\kappa_2$ 
for several cases of the particle hole excited states are shown 
in Table 3. 

From the fig.1, we know that 
 the breaking of the factorization of the S-matrix is 
the order of $|D_{1p1h}- D_{2p2h}|$ which is smaller than unity. 
We do not know 
whether this value of $|D_{1p1h}- D_{2p2h}|$ 
can be said to be  large or small. 
In any case, the S-matrix factorization is violated  
at the  elementary level.

Here, we again note that the S-matrix factorization for the particle 
particle scattering must hold. This is essentially the same as the vacuum 
case. It is interesting to observe that the energy of the 
vacuum state can be obtained  analytically, though the vacuum energy 
alone is not physically very meaningful. 

\end{enumerate}
\vspace{3cm}
\item{\Large Conclusions}

We have presented a numerical proof that the S-matrix factorization for the 
particle hole scattering in the massive Thirring model is not 
satisfied. This is mainly because the factorization of the S-matrix 
and the crossing symmetry do not commute with each other. 

It is always simpler to find a counter example of the violation 
of the S-matrix factorization than to prove the validity of the factorization. 
This is clear since we only have to find out any type of example 
which shows the violation of the factorization. This is indeed the point 
we have worked out in this paper. 
The present calculations present a convincing evidence that 
the S-matrix factorization for the particle hole scattering 
does not hold  exactly. 

Here, we want to examine the implication and  consequence of this result. 
As stressed in this paper, it is important to realize that the S-matrix 
factorization should hold for the particle particle scattering 
even for the massive Thirring model. This can be easily seen from 
the Bethe ansatz solution which is eq.(3.5). Indeed, if one looks at 
the vacuum to vacuum transition eq.(5.1), then one sees that the S-matrix 
is factorized into the product of the two body S-matrices. This is the 
consequence of the particle particle scattering. 
Clearly, the massless case 
must have this nice feature of the S-matrix factorization since, 
in this case, one can treat all the scattering processes as two species of 
fermion scattering without going to the particle hole scattering. 
Therefore, those models which are equivalent to the six vertex models 
should receive no effects from the present investigation.  

However, those models which are equivalent to the eight vertex models 
should be carefully treated if one is concerned with the 
bound state spectrum with S-matrix factorization ansatz. At least, 
we now know that the massive Thirring model does not have the S-matrix 
factorization property for the particle hole scattering. This result 
is consistent with recent calculations for the bound state of 
the massive Thirring model [15-17]. 
The Bethe ansatz as well as the light cone 
calculations show that there is only one bound state on the contrary 
to the prediction by the S-matrix factorization equation.

Here, we want to discuss the implication of the present result 
in connection with the Yang-Baxter equation. 
The Yang-Baxter equation is a matrix equation which is obtained 
by imposing certain conditions. The present study does not intend 
to check the  Yang-Baxter equation itself. Instead, 
we have only examined the factorization property of the S-matrix 
at the elementary level. Therefore, we have never questioned whether 
the requirement of the Yang-Baxter equation is physically reasonable 
or not. We think that the requirement of the Yang-Baxter equation 
is indeed reasonable if the S-matrix is factorizable at the elementary 
level. 

Now, the question is how we can interpret the results (the bound 
state spectrum) predicted by the Yang-Baxter equation when the S-matrix 
factorization is violated at the elementary level. 
In the massive Thirring / sine-Gordon model, the situation is 
now clear. Namely, the spectrum obtained by the Yang-Baxter 
equation with a violation of the S-matrix factorization 
at the elementary level is semiclassical. 

However, this point is only proved for the massive Thirring model. 
We do not know whether there might be some examples in which 
the Yang-Baxter equation gives exact spectrum even though the S-matrix 
factorization is violated at the elementary level.

\end{enumerate}

\vspace{1cm}
Acknowledgments: We  thank F. Lenz, M. Kato, C. Itoi, H. Seki and Takahashi 
 for stimulating discussions and comments. 

\vspace{2cm}


{\large REFERENCES }
\baselineskip = 8 mm

1. H.A. Bethe, Zeits f. Physik, {\bf 71} (1931),205

2. L.D. Faddeev and L.A. Takhtajan, Phys. Lett. {\bf 85A} (1981), 375

3. L.D. Faddeev and L.A. Takhtajan, J. Sov. Math. {\bf 24} (1984), 241

4. L.D. Faddeev, Int. J. Mod. Phys. {\bf 10} (1995), 1845

5. R.J. Baxter, Phys. Rev. Lett. {\bf 26}(1971), 832, 834

6. R.J. Baxter, Ann. Phys.  {\bf 70}(1972), 193, 323

7. R.J. Baxter, Ann. Phys. {\bf 76}(1973), 1,25,48

8. L.D. Fadeev and L.A. Takhtajan, Usp. Mat. Nauk {\bf 34} (1979), 13

9. M.F. Weiss and K.D. Schotte, Nucl. Phys. {\bf B225}(1983), 247

10. A. Luther, Phys. Rev. {\bf B14} (1976), 2153 

11. A.B.Zamolodchikov and A.B.Zamolodchikov, Ann. Phys. {\bf 120} (1979), 253 

12.  R. F. Dashen, B. Hasslacher and A. Neveu, 
Phys. Rev. {\bf D11} (1975), 3432 

13. H. Bergknoff and H.B. Thacker, Phys. Rev. Lett. {\bf 42} (1979), 135  

14. H.B. Thacker, Rev. Mod. Phys. {\bf 53} (1981), 253 

15. T. Fujita, Y. Sekiguchi and K. Yamamoto, Ann. Phys. {\bf 255} 
(1997), 204

16. T. Fujita and A. Ogura,  Prog. Theor. Phys. {\bf 89} (1993), 23 

17. A. Ogura, T. Tomachi and T. Fujita, Ann. Phys. (N.Y.) 
{\bf 237} (1995), 12

18. M. Cavicchi, Int. J. Mod. Phys. {\bf A10} (1995), 167 

19. W. Thirring, Ann. Phys. (N.Y) {\bf 3} (1958), 91 

20. M. Gomes and J.H. Lowenstein, Nucl. Phys. {\bf B45} (1972), 252

\vspace{3cm}

{\Large Figure captions : }

\vspace{0.5cm}

\begin{list}{}{}
\item[Fig.1a]: 
 We show the calculated values of $D_{1p1h}$ and 
$D_{2p2h}$ as the function of $\rho_0^{-1}$. This is the excited state with  
$n_{i_0}=1$, $n_{i_1}=1$ and $n_{i_2}=-1$.  The solid line is drawn 
with eqs.(5.24) with the parameters in Table 3. 

\item[Fig.1b]: The same as Fig.1a. This is the excited state with  
$n_{i_0}=2$, $n_{i_1}=1$ and $n_{i_2}=2$.

\item[Fig.1c]: The same as Fig.1a. This is the excited state with  
$n_{i_0}=2$, $n_{i_1}=1$ and $n_{i_2}=-2$.

\end{list} 
\newpage

\begin{center}
\underline{Table 1} \\
\ \ \\
Rapidities \\
\ \ \\
\begin{tabular}{|c|c||c|c|c|}
\hline
$ $ & $ $ & $ $ & $ $ & $ $ \\
\   & $N$ 
& $\alpha_{n_i}$ & $\alpha_{n_i}^{\dagger}$ & $\alpha_{n_i}^{\ddagger}$  \\
$ $ & $ $ & $ $ & $ $ & $ $ \\ \hline
$n_{i}=5$ & 
\begin{tabular}{c} 
101 \\ 401 \\ 1601 \\ 6401 \\ 12801
\end{tabular} & 
\begin{tabular}{r@{\extracolsep{0pt}.}l}
 0 & 859 \\
 0 & 223 \\
 0 & 0559 \\
 0 & 0140 \\
 0 & 0070      
\end{tabular} & 
\begin{tabular}{r@{\extracolsep{0pt}.}l}
 0 & 801 \\
 0 & 202 \\
 0 & 0504 \\
 0 & 0126 \\
 0 & 0063      
\end{tabular} & 
\begin{tabular}{r@{\extracolsep{0pt}.}l}
 1 & 142 \\
 0 & 247 \\
 0 & 0574 \\
 0 & 0141 \\
 0 & 0070  
\end{tabular}  \\
\hline
$n_{i}=10$ & 
\begin{tabular}{c} 
101 \\ 401 \\ 1601 \\ 6401 \\ 12801 
\end{tabular} & 
\begin{tabular}{r@{\extracolsep{0pt}.}l}
 1 & 578 \\
 0 & 443 \\
 0 & 1118 \\
 0 & 0280 \\
 0 & 0140  
\end{tabular} & 
\begin{tabular}{r@{\extracolsep{0pt}.}l}
 1 & 546 \\
 0 & 424 \\
 0 & 1064 \\
 0 & 0266 \\
 0 & 0133  
\end{tabular} & 
\begin{tabular}{r@{\extracolsep{0pt}.}l}
 1 & 863 \\
 0 & 486 \\
 0 & 1147 \\
 0 & 0282 \\
 0 & 0140  
\end{tabular} \\
\hline
$n_{i}=15$ & 
\begin{tabular}{c} 
101 \\ 401 \\ 1601 \\ 6401 \\ 12801 
\end{tabular} & 
\begin{tabular}{r@{\extracolsep{0pt}.}l}
 2 & 147 \\
 0 & 656 \\
 0 & 1676 \\
 0 & 0420 \\
 0 & 0210  
\end{tabular} & 
\begin{tabular}{r@{\extracolsep{0pt}.}l}
 2 & 128 \\
 0 & 640 \\
 0 & 1623 \\
 0 & 0406 \\
 0 & 0203  
\end{tabular} & 
\begin{tabular}{r@{\extracolsep{0pt}.}l}
 2 & 390 \\
 0 & 713 \\
 0 & 1718 \\
 0 & 0422 \\
 0 & 0210  
\end{tabular} \\
\hline
$n_{i}=20$ & 
\begin{tabular}{c} 
101 \\ 401 \\ 1601 \\ 6401 \\ 12801 
\end{tabular} & 
\begin{tabular}{r@{\extracolsep{0pt}.}l}
 2 & 603 \\
 0 & 862 \\
 0 & 2232 \\
 0 & 0559 \\
 0 & 0280  
\end{tabular} & 
\begin{tabular}{r@{\extracolsep{0pt}.}l}
 2 & 592 \\
 0 & 848 \\
 0 & 2181 \\
 0 & 0546 \\
 0 & 0273  
\end{tabular} & 
\begin{tabular}{r@{\extracolsep{0pt}.}l}
 2 & 806 \\
 0 & 926 \\
 0 & 2288 \\
 0 & 0563 \\
 0 & 0281  
\end{tabular} \\
\hline
\end{tabular}
\\
\vspace*{2cm}  

\begin{minipage}{13cm}
We plot the rapidity values for several cases of $n_i$ for  
the vacuum, $1p-1h$ state with $n_{i_0}=0$ and $2p-2h$ state 
with $n_{i_1}=1$, $n_{i_2}=-1$.  
The density $\rho_0$ is fixed to $\rho_0=20$. 
\end{minipage}
\end{center}

\vspace{3cm}

\newpage

\begin{center}
\underline{Table 2} \\
\ \ \\
$\displaystyle{ (g_0=6.28) }$  \\
\ \ \\
\begin{tabular}{|c|c||c|c||c|c|}
\hline
$ $ & $ $ & $ $ & $ $ & $ $ & $ $ \\
   & $N$ & $D_{1p1h}(n_{i_0})$ &  $E_{1p1h}(n_{i_0})$ 
   & $D_{2p2h}(n_{i_1},n_{i_2})$ &  $E_{2p2h}(n_{i_1},n_{i_2}) $ \\
$ $ & $ $ & $ $ & $ $ & $ $ & $ $ \\   
\hline
\hline
$\rho_0 =20$ & 
\begin{tabular}{c} 
101 \\ 401 \\ 1601 \\ 6401 \\ 12801
\end{tabular} & 
\begin{tabular}{r@{\extracolsep{0pt}.}l}
0 & 277 \\
0 & 261 \\
0 & 257 \\
0 & 256 \\
0 & 256      
\end{tabular} & 
\begin{tabular}{r@{\extracolsep{0pt}.}l}
 1 & 00 $\times 10^{-2}$ \\
 2 & 40 $\times 10^{-3}$ \\
 5 & 94 $\times 10^{-4}$ \\
 1 & 48 $\times 10^{-4}$ \\
 7 & 39 $\times 10^{-5}$ 
\end{tabular} &
\begin{tabular}{r@{\extracolsep{0pt}.}l}
 0 & 767 \\
 0 & 764 \\
 0 & 762 \\
 0 & 763 \\
 0 & 762     
\end{tabular} & 
\begin{tabular}{r@{\extracolsep{0pt}.}l}
 7 & 31 $\times 10^{-3}$ \\
 1 & 73 $\times 10^{-3}$ \\
 4 & 25 $\times 10^{-4}$ \\
 1 & 06 $\times 10^{-4}$ \\
 5 & 28 $\times 10^{-5}$ 
\end{tabular}  \\
\hline
\end{tabular}
\\
\vspace*{2cm}  

\begin{minipage}{13cm}
We plot the values of $D_{1p1h}$, $E_{1p1h}$, $D_{2p2h}$ and $E_{2p2h}$ 
for several cases of the number of the particles $N$ for the lowest 
scattering states ( $n_{i_0}=1$, $n_{i_1}=1$, $n_{i_2}=-1$. ). 
 The density $\rho_0$ is fixed to $\rho_0=20$.  

\end{minipage}
\end{center}


\newpage
\begin{center}
\underline{Table 3a} \\
\ \ \\

  $ D_{1p1h}(n_{i_0}) $ \ \ $(g_0=6.28)$   \\
\ \ \\
\begin{tabular}{|l||r@{\extracolsep{0pt}.}l|r@{\extracolsep{0pt}.}l| }
\hline
 & \multicolumn{2}{c|}{ } & 
 \multicolumn{2}{c|}{ }   \\ 
 & \multicolumn{2}{c|}{$\displaystyle{ C^{(1)} }$} & 
 \multicolumn{2}{c|}{$\displaystyle{ D^{(1)} }$}   \\
 & \multicolumn{2}{c|}{ } & 
 \multicolumn{2}{c|}{ }   \\ 
\hline
 & \multicolumn{2}{c|}{ } & 
 \multicolumn{2}{c|}{ }  \\ 
$ n_{i_0}=1 $ & 
 $-$0 & 359(16) &
 0 & 277(1)   \\
 & \multicolumn{2}{c|}{ } & 
 \multicolumn{2}{c|}{ }   \\ 
$  n_{i_0}=-1 $ &  
 0 & 359(16) &
 $-$0 & 277(1)  \\
 & \multicolumn{2}{c|}{ } & 
 \multicolumn{2}{c|}{ }   \\ 
$  n_{i_0}=2 $ &  
 $-$0 & 329(14) &
 0 & 275(1)  \\
 & \multicolumn{2}{c|}{ } & 
 \multicolumn{2}{c|}{ }   \\ 
$  n_{i_0}=-2 $ &  
 0 & 329(14) &
 $-$0 & 275(1)  \\
 & \multicolumn{2}{c|}{ } & 
 \multicolumn{2}{c|}{ }   \\ 
\hline
\end{tabular}
\\
\vspace*{1cm}  
\begin{minipage}{13cm}
\begin{center}

We plot the values of the parameters $C^{(1)}$, $D^{(1)}$. 
The numbers in parenthesis indicate  error bars in the fitting. 
\end{center}
  
\end{minipage}
\newpage

\underline{Table 3b} \\
\ \ \\

 $ D_{2p2h}(n_{i_1},n_{i_2}) $ \ \ $(g_0=6.28)$ \\
\ \ \\
\begin{tabular}{|l||r@{\extracolsep{0pt}.}l|r@{\extracolsep{0pt}.}l|
r@{\extracolsep{0pt}.}l|}
\hline
 & \multicolumn{2}{c|}{ } & 
 \multicolumn{2}{c|}{ } & 
 \multicolumn{2}{c|}{ }  \\ 
 & \multicolumn{2}{c|}{$\displaystyle{ C^{(2)} }$} & 
 \multicolumn{2}{c|}{$\displaystyle{ D^{(2)} }$} &
 \multicolumn{2}{c|}{$\displaystyle{ \kappa_2 }$}  \\
 & \multicolumn{2}{c|}{ } & 
 \multicolumn{2}{c|}{ } & 
 \multicolumn{2}{c|}{ }  \\ 
\hline
 & \multicolumn{2}{c|}{ } & 
 \multicolumn{2}{c|}{ } & 
 \multicolumn{2}{c|}{ }  \\ 
$  n_{i_1}=1,n_{i_2}=-1 $ & 
 0 & 404(8) &
 0 & 626(2) &
 22 & 0  \\
 & \multicolumn{2}{c|}{ } & 
 \multicolumn{2}{c|}{ } & 
 \multicolumn{2}{c|}{ }  \\ 
$   n_{i_1}=1,n_{i_2}=2  $ &  
 0 & 115(1) &
 0 & 454(1) &
 12 & 0  \\
 & \multicolumn{2}{c|}{ } & 
 \multicolumn{2}{c|}{ } & 
 \multicolumn{2}{c|}{ }  \\ 
$   n_{i_1}=1,n_{i_2}=-2   $ & 
 0 & 388(5) &
 0 & 603(1) &
 18 & 0  \\
 & \multicolumn{2}{c|}{ } & 
 \multicolumn{2}{c|}{ } & 
 \multicolumn{2}{c|}{ }  \\ 
$  n_{i_1}=-1,n_{i_2}=1 $ & 
 $-$0 & 404(8) &
 $-$0 & 626(1) &
 22 & 0  \\
 & \multicolumn{2}{c|}{ } & 
 \multicolumn{2}{c|}{ } & 
 \multicolumn{2}{c|}{ }  \\ 
$   n_{i_1}=2,n_{i_2}=1  $ &  
 0 & 115(1) &
 0 & 454(1) &
 12 & 0  \\
 & \multicolumn{2}{c|}{ } & 
 \multicolumn{2}{c|}{ } & 
 \multicolumn{2}{c|}{ }  \\ 
$   n_{i_1}=-2,n_{i_2}=1   $ & 
 $-$0 & 388(5) &
 $-$0 & 603(1) &
 18 & 0  \\
 & \multicolumn{2}{c|}{ } & 
 \multicolumn{2}{c|}{ } & 
 \multicolumn{2}{c|}{ }  \\ 
\hline
\end{tabular}
\\
\vspace*{1cm}  

\begin{minipage}{13cm}
\begin{center}
We plot the values of the parameters $C^{(2)}$, $D^{(2)}$ and 
$\kappa_2$. 
The numbers in parenthesis indicate error bars in the fitting. 
  
\end{center}
\end{minipage}

\end{center}

\end{document}